\documentclass[a4paper]{jpconf}
\usepackage{graphicx}
\usepackage{dcolumn}
\usepackage{bm}% bold math
\usepackage{color}
\usepackage{ulem}
\usepackage{amsmath}

\begin{document}
\title{Quantum spin Hall effect in a two-orbital model on a honeycomb lattice}

\author{Satoru Hayami$^1$, Hiroaki Kusunose$^2$, and Yukitoshi Motome$^1$}

\address{$^1$ Department of Applied Physics, University of Tokyo, Tokyo 113-8656, Japan}
\address{$^2$ Department of Physics, Ehime University, Matsuyama 790-8577, Japan
}

\ead{hayami@aion.t.u-tokyo.ac.jp}

\begin{abstract}
The spin Hall effect is investigated in a two-orbital tight-binding model on a honeycomb lattice. 
We show that the model exhibits three topologically-different 
insulating phases at half filling, which are distinguished by different quantized values of the spin Hall conductivity. 
We analytically determine the phase boundaries, where the valence and conduction bands touch with each other with forming the Dirac nodes at the Fermi level. 
The results are discussed in terms of the effective antisymmetric spin-orbit coupling. 
The relation to the Kane-Mele model and 
implications for a magnetoelectric effect are also discussed. 
\end{abstract}

\section{Introduction}
The spin-orbit coupling has drawn much interest in condensed matter physics since it leads to various fascinating phenomena in spin-charge-orbital coupled systems, such as the magetoelectric effect~\cite{curie1894symetrie,dzyaloshinskii1960magneto,kimura2003magnetic,wang2003epitaxial,KhomskiiPhysics.2.20}, the spin Hall effect~\cite{hirsch1999spin,murakami2003dissipationless,sinova2004universal}, and the noncentrosymmetric superconductivity~\cite{Bauer_PhysRevLett.92.027003,Bauer_Sigrist201201}. 
Among them, the quantum spin Hall effect in topological insulators has been extensively studied both theoretically and experimentally~\cite{Haldane_PhysRevLett.61.2015,Kane_PhysRevLett.95.146802,Kane_PhysRevLett.95.226801,Murakami_PhysRevLett.97.236805,bernevig2006quantum,Hasan_RevModPhys.82.3045}. 
There, the spin Hall conductivity is quantized at a certain value, which is determined by the topological number distinguishing topologically-different states. 

A key concept in understanding of such phenomena is the antisymmetric spin-orbit coupling. It is written in a general form in the 
momentum representation: 
\begin{align}
\label{eq:Ham_ASOC}
\mathcal{H}_{{\rm ASOC}} = \alpha \sum_{\bm{k}} \bm{g} (\bm{k}) \cdot \bm{s} (\bm{k}), 
\end{align}
where $\alpha$ represents the magnitude of the antisymmetric spin-orbit coupling, and $\bm{g}(\bm{k})$ is called the $g$ vector; 
$\bm{s}(\bm{k})$ is the spin operator defined in the momentum space as 
$\bm{s}(\bm{k}) = (1/2) \sum_{\sigma \sigma'} c^{\dagger}_{\bm{k} \sigma} \bm{\sigma}_{\sigma \sigma'} c_{\bm{k} \sigma}$
where $c_{\bm{k}\sigma}^{\dagger}$ ($c_{\bm{k}\sigma}$) is a creation (annihilation) operator at wave vector $\bm{k}$ and spin $\sigma$, and $\bm{\sigma}$ is the vector of Pauli matrices. 
In Eq.~(\ref{eq:Ham_ASOC}),  
$\bm{g}(\bm{k})$ is antisymmetric with respect to $\bm{k}$, and 
the direction of the $g$ vector is determined by the symmetry of the crystal. 
For example, when the mirror symmetry along the $z$ direction is broken, the so-called Rashba-type antisymmetric spin-orbit coupling exists, whose $g$ vector is given by 
$\bm{g}(\bm{k}) = (k_y, -k_x,0)$~\cite{rashba1960properties,bychkov1984oscillatory}.

In the present study, we investigate the role of a ``site-dependent" antisymmetric spin-orbit coupling. 
Namely, considering a lattice structure in which the spatial-inversion (parity) symmetry is preserved globally but broken intrinsically at each site, we discuss the effect of a hidden $g$ vector in the site-dependent form. 
The spatially-modulated antisymmetric spin-orbit coupling has recently been studied in multilayer superconductors~\cite{Maruyama_doi:10.1143/JPSJ.81.034702,Yoshida_PhysRevB.86.134514,Yoshida_doi:10.7566/JPSJ.82.074714}.
Here, we study such effect in a generic two-orbital tight-binding model on a honeycomb lattice. 
The model was studied by the authors with focusing on the symmetry-broken states by 
electron correlation~\cite{Hayami_PhysRevB.90.081115}, but here we concentrate on the paramagnetic state in the noninteracting limit. 
We note that related $p$-orbital models were also discussed in the context of cold atoms~\cite{Wu_PhysRevLett.99.070401}. 
We find that our model at half filling exhibits three topologically-different insulating states showing the quantum spin Hall effect. 
We discuss their topological nature in terms of the band structure and the 
hidden antisymmetric spin-orbit coupling.

\section{Model}

\begin{figure}[htb!]
\begin{minipage}{18pc}
\centering
\includegraphics[width=18pc]{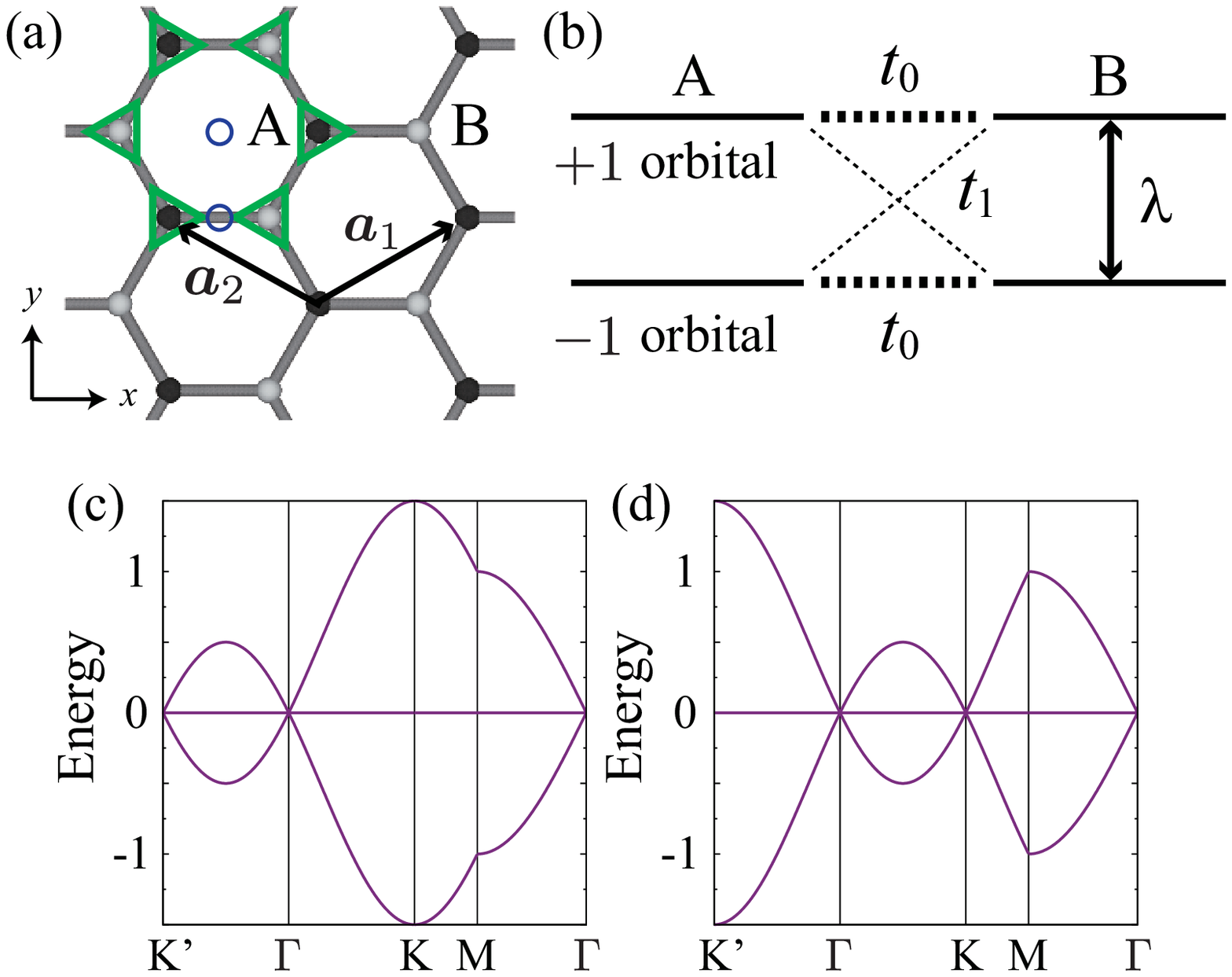} 
\caption{
\label{Fig:gamma}
(a) Schematic picture of a honeycomb lattice; 
the primitive translation vectors are $\bm{a}_1 = (\sqrt{3}/2, 1/2)$ and $\bm{a}_2 = (-\sqrt{3}/2, 1/2)$. 
Open circles (triangles) indicate the inversion centers (the parity-broken sites). 
(b) Schematic picture of the energy levels of the two-band model Hamiltonian in Eq.~(\ref{Eq:H0}). 
(c) and (d) Band dispersions when we consider only $\gamma_{+1 \bm{k}}$ and 
$\gamma_{-1 \bm{k}}$ in Eq.~(\ref{Eq:H0}) with $t_1 = 0.5$, respectively, at $t_0=\lambda=0$. 
}
\end{minipage}\hspace{2pc}
\begin{minipage}{18pc}
\includegraphics[width=18pc]{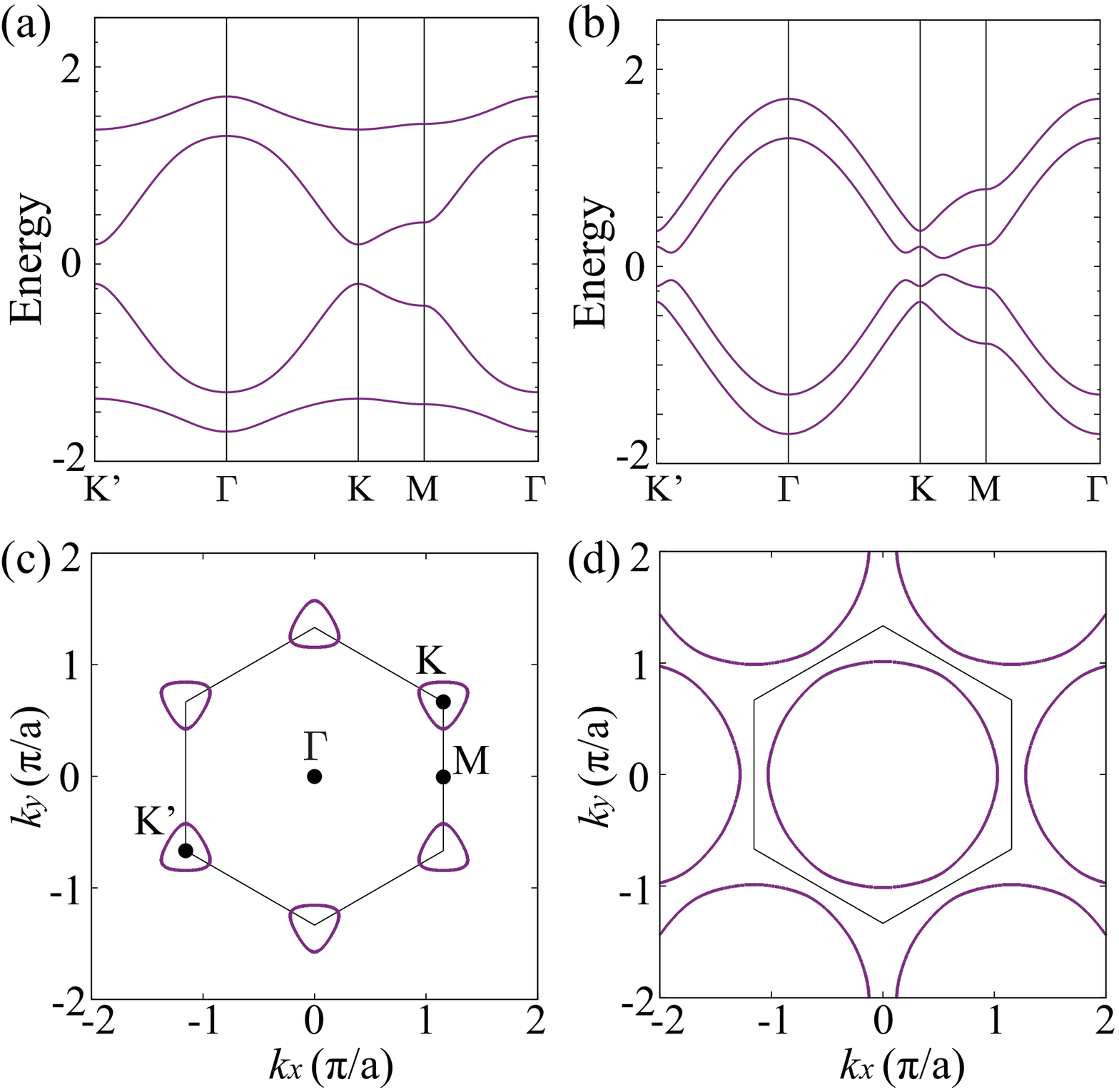} 
\caption{
\label{Fig:paramagnetic_band}
Electronic band structures of the model in Eq.~(\ref{Eq:H0}) at $t_0=0.5$ and $\lambda=0.4$: (a) $t_1 =0.45$ and (b) $t_1 = 0.1$, shown along the symmetric lines in the first Brillouin zone. 
(c) and (d) Energy contours slightly below the Fermi level at half filling 
($E=-0.3$) corresponding to (a) and (b), respectively. 
The hexagon represents the first Brillouin zone. 
}
\end{minipage}
\end{figure}

We consider a minimal multi-orbital model on a honeycomb lattice [see Fig.~\ref{Fig:gamma}(a)], which was recently proposed by the authors~\cite{Hayami_PhysRevB.90.081115}. 
Implicitly assuming a large crystalline electric field, we consider only a pair of $d$ orbitals with the angular momenta $m=\pm1$. 
A generalization for the $m=\pm2$ case is straightforward. 
Then, the tight-binding Hamiltonian is written as 
\begin{align}
\label{Eq:H0}
\mathcal{H} 
= - \sum_{\bm{k}m\sigma} 
(  t_0 \gamma_{0\bm{k}} c_{{\rm A}\bm{k}
m\sigma}^{\dagger} c_{{\rm B}\bm{k} 
m\sigma} +  t_1 
\gamma_{m\bm{k}}  c_{{\rm A}\bm{k}
m\sigma}^{\dagger} c_{{\rm B}\bm{k} - 
m\sigma} + {\rm H.c.}) +\frac{\lambda}{2}\sum_{s\bm{k}m\sigma} 
c_{s \bm{k} 
m\sigma}^{\dagger} (m\sigma) 
c_{s \bm{k} 
m\sigma},  
\end{align}
where $c_{s \bm{k} m\sigma}^{\dagger}$ ($c_{s \bm{k} m\sigma}$) is the creation (annihilation) operator for sublattice $s=$ A 
or B,  
wave number $\bm{k}$, orbital $m=\pm 1$, and spin $\sigma = \uparrow$ or $\downarrow$. 
The first and second terms represent the intra- and inter-orbital 
hoppings between nearest-neighbor sites, respectively. 
The third term in Eq.~(\ref{Eq:H0}) represents the atomic spin-orbit coupling, 
which has a nonzero matrix element only for the $z$ component as $m=\pm1$. 
The schematic picture of the energy levels of the Hamiltonian in Eq.~(\ref{Eq:H0}) is shown in Fig.~\ref{Fig:gamma}(b). 

The $\bm{k}$ dependence 
in the hopping terms is given by 
\begin{align} 
\gamma_{ n\bm{k}} &= e^{i \bm{k}\cdot \bm{\eta}_1 } +  \omega^{ 
-2 n} e^{{\rm i} \bm{k}\cdot \bm{\eta}_2 }+ \omega^{2n} 
e^{{\rm i} \bm{k}\cdot \bm{\eta}_3 }
=\gamma_{-n,-\bm{k}}^{*}, 
\label{eq:gamma}
\end{align}
where $\omega = e^{2\pi {\rm i}/3}$;
$\bm{\eta}_1 = (\bm{a}_1 - \bm{a}_2)/3$, $\bm{\eta}_2 = (\bm{a}_1 + 2 \bm{a}_2)/3$, and $\bm{\eta}_3 =-(2 \bm{a}_1 + \bm{a}_2)/3$ [$\bm{a}_1 = (\sqrt{3}/2,1/2)$ and $\bm{a}_2= (-\sqrt{3}/2)$ are primitive translational vectors; see Fig.~\ref{Fig:gamma}(a)]. 
The additional phase factors in Eq.~(\ref{eq:gamma}) come from transfers between orbitals with the different angular momenta. 
Note that $\gamma_{m\bm{k}}$ with $m=\pm1$ bring about the antisymmetry with respect to $\bm{k}$. 
For instance, Figs.~\ref{Fig:gamma}(c) and \ref{Fig:gamma}(d) show the antisymmetric behavior of $\gamma_{\pm 1\bm{k}}$ at $t_1=0.5$. 
The contrasting asymmetry takes place between the K and K' points. 
This indicates that the local asymmetric behavior is hidden under global inversion symmetry in the honeycomb lattice. 
The asymmetry in $\bm{k}$ plays an important role in the emergence of spin Hall effect discussed below.

\section{Result and Discussion}
\label{sec:Paramagnetic State}

Figures~\ref{Fig:paramagnetic_band}(a) and \ref{Fig:paramagnetic_band}(b) show the band structures in the model in Eq.~(\ref{Eq:H0}) at $t_0=0.5$ and $\lambda=0.4$: 
(a) $t_1=0.45$ and (b) $t_1=0.1$. 
As both spatial-inversion and time-reversal symmetries are preserved, each band is doubly degenerate. 
The energy gaps at commensurate fillings [1/4, half, 3/4 fillings in (a), and half filling in (b)] originate from both the atomic spin-orbit coupling and inter-orbital hopping. 
At half filling, the energy gap opens at the K and K' points for large $t_1$ as shown in Fig.~\ref{Fig:paramagnetic_band}(a), whereas it opens at the incommensurate points for small $t_1$ as shown in Fig.~\ref{Fig:paramagnetic_band}(b). 
Reflecting the difference in the band structures, the energy contours around the Fermi level at half filling are significantly dependent on the values of $t_1$. 
Figures~\ref{Fig:paramagnetic_band}(c) and \ref{Fig:paramagnetic_band}(d) show the energy contours at the energy $E=-0.3$. 
They are topologically different: the energy contours are the small pockets around the K and K' points in Fig.~\ref{Fig:paramagnetic_band}(a), while the large circle around the $\Gamma$ pint in Fig.~\ref{Fig:paramagnetic_band}(b). 
Such differences reflect the distinct topological nature as detailed below.

\begin{figure}[htb!]
\centering
\includegraphics[width=1.0 \hsize]{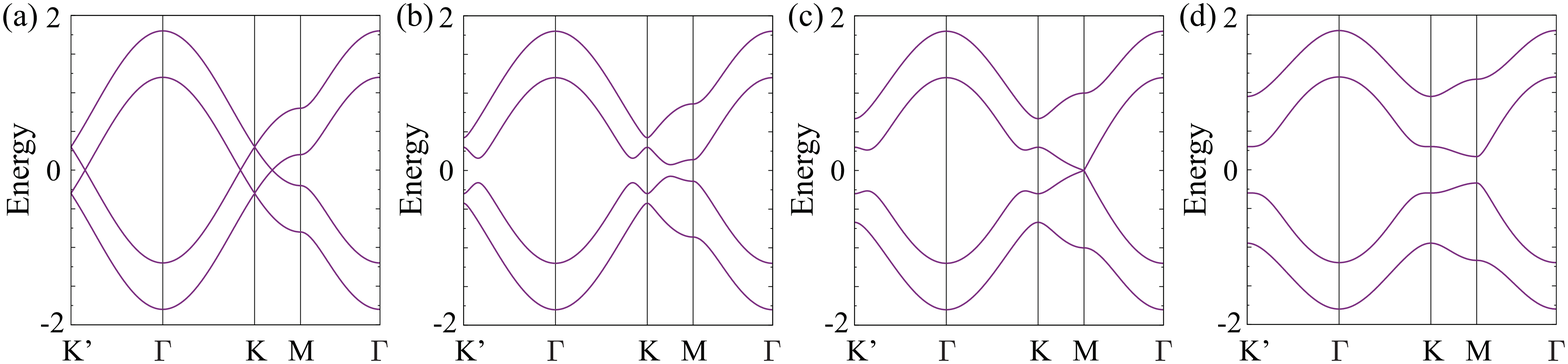} 
\caption{
Electronic band structures of the model in Eq.~(\ref{Eq:H0}) at $t_0=0.5$ and $\lambda=0.6$: (a) $t_1 =0$, (b) $t_1 = 0.1$, (c) $t_1 = 0.2$, and (d) $t_1=0.3$. 
In (c), the valence and conduction bands touch with each other at the M point. 
}
\label{Fig:paramagnetic_band_yline}
\end{figure}

\begin{figure}[htb!]
\centering
\includegraphics[width=1.0 \hsize]{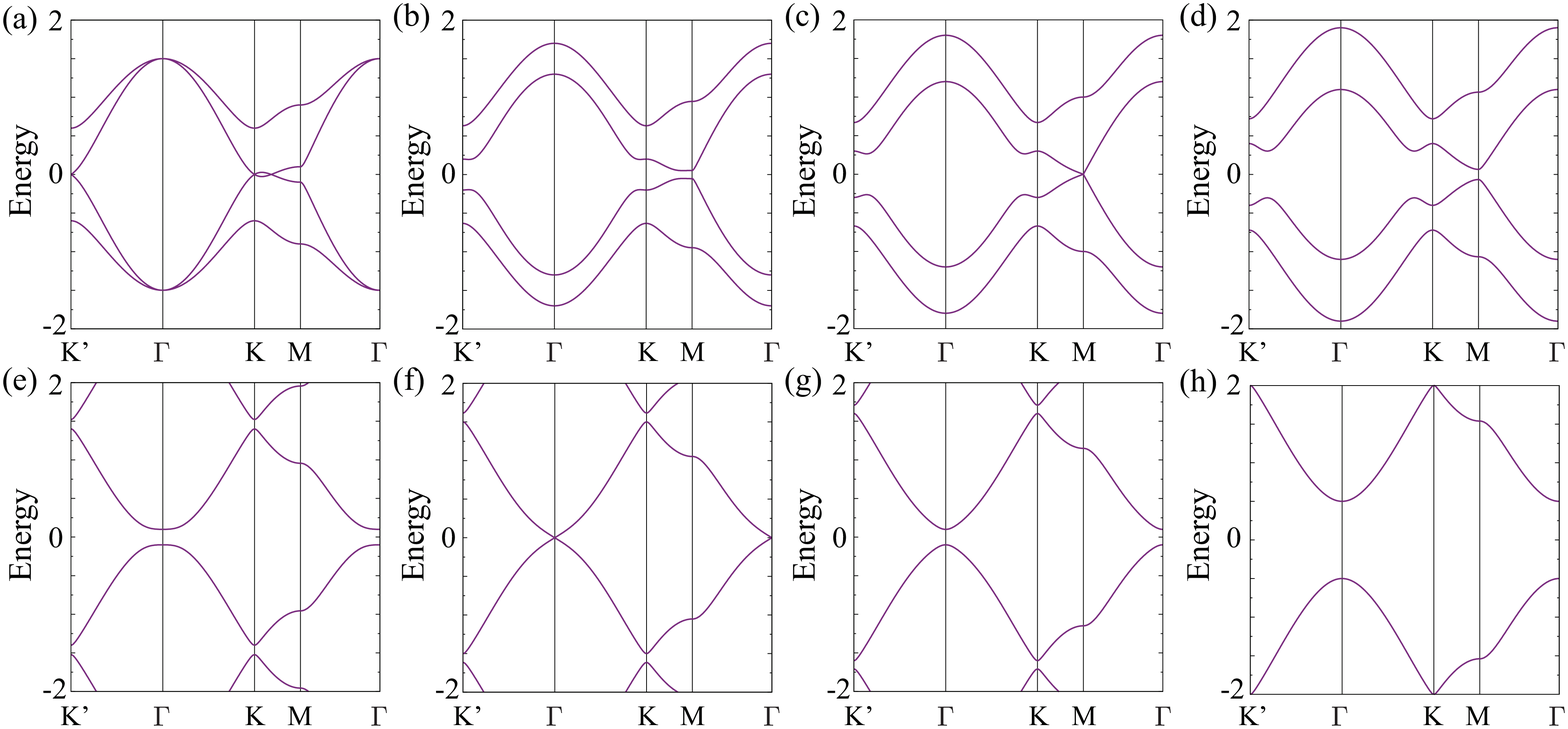} 
\caption{
Electronic band structures of the model in Eq.~(\ref{Eq:H0}) at $t_0=0.5$ and $t_1=0.2$: (a) $\lambda =0$, (b) $\lambda = 0.4$, (c) $\lambda = 0.6$, (d) $\lambda=0.8$, (e) $\lambda = 2.8$, (f) $\lambda=3$, (g) $\lambda=3.2$, and (h) $\lambda=4.0$. 
In (c) and (f), the valence and conduction bands at half filling touch with each other at the M nd $\Gamma$ points, respectively. 
}
\label{Fig:paramagnetic_band_xline}
\end{figure}

The topological nature of insulators may change 
when the valence and conduction bands touch with each other. 
The energy eigenvalues of the Hamiltonian in Eq.~(\ref{Eq:H0}) are given by 
\begin{align}
&\varepsilon (\bm{k}) = \pm \frac{1}{2}
\biggl\{
\lambda^2 + 4 |\tilde{\gamma}_{0 \bm{k}}|^2  + 2 |\tilde{\gamma}_{-1 \bm{k}}|^2 
+ 2  |\tilde{\gamma}_{1 \bm{k}}|^2 \biggr. \nonumber \\
&\biggl. \pm 2 \sqrt{ (|\tilde{\gamma}_{-1 \bm{k}}|^2-|\tilde{\gamma}_{1 \bm{k}}|^2)^2 
+   
8{\rm Re} \left[
\tilde{\gamma}_{-1 \bm{k}}\tilde{\gamma}_{1 \bm{k}} (\tilde{\gamma}_{0 \bm{k}}^*)^2 \right]
+4 |\tilde{\gamma}_{0 \bm{k}}|^2  (|\tilde{\gamma}_{-1 \bm{k}}|^2
+|\tilde{\gamma}_{1 \bm{k}}|^2)
+4  |\tilde{\gamma}_{0 \bm{k}}|^2  \lambda^2
}
\biggr\}^{\frac{1}{2}},  
\end{align}
where $\tilde{\gamma}_{0  \bm{k}} = t_0 \gamma_{0 \bm{k}}$ and $\tilde{\gamma}_{\pm1   \bm{k}}=t_1\gamma_{\pm1 \bm{k}}$. 
Especially, the energy eigenvalues at the symmetric points, $\Gamma$, K, K', and M, are given by 
\begin{align}
\label{eq:varepsilon gamma}
\varepsilon_{\Gamma} &= \pm 3t_0 \pm \frac{1}{2}\lambda,  \\
\label{eq:varepsilon K}
\varepsilon_{{\rm K}} &= \varepsilon_{{\rm K'}} = \pm \frac{1}{2} \lambda, \ \pm \frac{1}{2}\sqrt{\lambda^2+(6 t_1)^2},  \\
\label{eq:varepsilon M}
\varepsilon_{{\rm M}} &= \pm \frac{1}{2} 
\left\{ \lambda^2 + (2 t_0)^2 + (4 t_1)^2 \pm 2\sqrt{\lambda^2(2 t_0)^2 + (2 t_0)^2 (4 t_1)^2}
\right\}^{\frac{1}{2}}, 
\end{align}
respectively. Note that all the energies are doubly degenerate. 
From Eqs.~(\ref{eq:varepsilon gamma})-(\ref{eq:varepsilon M}), the band touching occurs when the following conditions 
are satisfied:  
\begin{align}
\label{eq:band_touching1}
4t_1 
= \sqrt{(2t_0)^2-\lambda^2} 
\quad &({\rm gap \ closes \ at \ the \ M \ point}), \\
\label{eq:band_touching2}
\lambda 
= 6 t_0 
\quad &({\rm gap \ closes \ at \ the} \ \Gamma  \ {\rm point}).
\end{align}
Such band touching is demonstrated while changing $t_1$ and $\lambda$ 
in Figs.~\ref{Fig:paramagnetic_band_yline} and \ref{Fig:paramagnetic_band_xline}, respectively. 
The results suggest the topological changes of the paramagnetic insulating states at half filling.

\begin{figure}[htb!]
\centering
\includegraphics[width=1.0 \hsize]{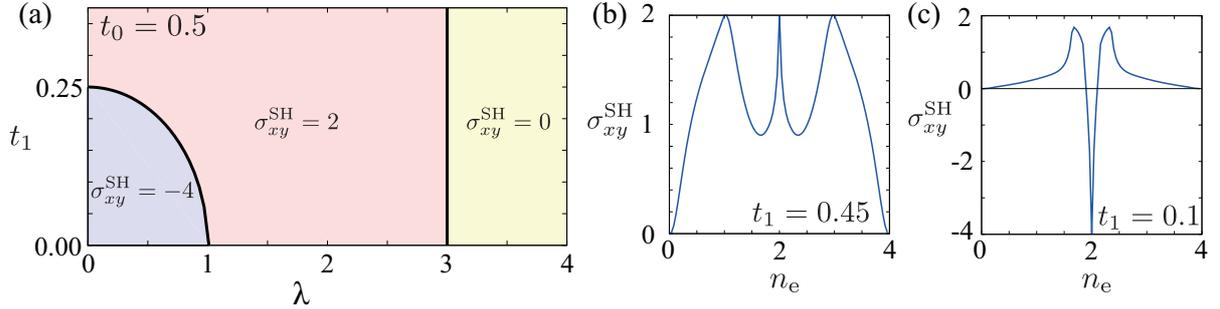}
\caption{
\label{Fig:paramagnetic_Hall}
(a) The phase diagram for the paramagnetic state at half filling as a function of $\lambda$ and $t_1$.  
The different phases are characterized by the different values of the quantized spin 
Hall conductivity, $\sigma_{xy}^{{\rm SH}}$. 
(b) and (c) Filling dependences of the spin Hall conductivity. 
The data are calculated at $t_0=0.5$ and $\lambda=0.4$: (a) $t_1=0.45$ and (b) $t_1=0.1$, which are the same as in Figs.~\ref{Fig:paramagnetic_band}(a) and \ref{Fig:paramagnetic_band}(b), respectively. 
}
\end{figure}

We here identify such topological changes by the spin Hall effect. 
We compute the spin Hall conductivity by using the Kubo formula as 
\begin{align}
\label{eq:sigmaxy_spin}
\sigma_{xy}^{{\rm SH}} =-\frac{e}{2\hbar} \frac{1}{{\rm i} V_0} \sum_{\alpha \beta \bm{k}} \frac{f(\varepsilon_{\beta \bm{k}})-f(\varepsilon_{\alpha \bm{k}})}{\varepsilon_{\beta \bm{k}}-\varepsilon_{\alpha \bm{k}}} \frac{J^{(s) \beta \alpha
}_{x,\bm{k}} J^{ \alpha \beta}_{y,\bm{k}}}{\varepsilon_{\beta \bm{k}}-\varepsilon_{\alpha \bm{k}} + {\rm i}\delta}, 
\end{align}
where $V_0$ is the system volume, $f(\varepsilon)$ is the Fermi distribution function, and $\varepsilon_{\alpha \bm{k}}$ and $| \alpha \bm{k} \rangle$ are the eigenvalue and eigenstate of the Hamiltonian for the paramagnetic state. 
Here, $J_{\nu, \bm{k} 
}^{(s) \alpha \beta} = \langle \alpha \bm{k} | J_{\nu}^{({\rm s})}| \beta \bm{k} \rangle$ is the matrix element of the spin current operator $J_\nu^{({s})}$ which is defined by $J_\nu^{({ s})}= \frac{1}{2}\{\sigma_z, J_{\nu}\}$ in the $\nu$ direction ($\sigma_z$ is 
the $z$-component spin operator, $J_{\nu}$ is the current operator, and $\{ \cdots \}$ is an anticommutator); meanwhile $J_{\nu, \bm{k}}^{\alpha \beta} = \langle \alpha \bm{k} | J_{\nu}| \beta \bm{k} \rangle$. 
We set $-e/2\hbar=1$.
Thus, $\sigma_{xy}^{{\rm SH}}$ represents the coefficient for the spin current in the $x$ direction induced by the electric field in the $y$ direction [see Fig.~\ref{Fig:gamma}(a)]. 
We take temperature $T=0.001$ and the damping factor $\delta=0.001$. 

As a result, we find that the spin Hall conductivity at half filling is quantized at a different integer value depending on the parameters, reflecting the topologically-different nature in each insulating region. 
Figure~\ref{Fig:paramagnetic_Hall}(a) shows the phase diagram at half filling and $t_0=0.5$, 
labeled by the values of $\sigma_{xy}^{{\rm SH}}$; 
the three regions possess different quantized values of 
$\sigma_{xy}^{\rm SH}$, $-4$, $0$, and $2$. 
The phase boundaries are given by Eqs.~(\ref{eq:band_touching1}) and (\ref{eq:band_touching2}), where 
the system shows the Dirac nodes at the Fermi level, as shown in Figs.~\ref{Fig:paramagnetic_band_yline}(c), \ref{Fig:paramagnetic_band_xline}(c) and \ref{Fig:paramagnetic_band_xline}(f). 

Figures~\ref{Fig:paramagnetic_Hall}(b) and \ref{Fig:paramagnetic_Hall}(c) show 
$\sigma_{xy}^{{\rm SH}}$ as a function of the electron density $n_{{\rm e}} =(1/2 N_{\bm k}) \sum_{s\bm{k} m \sigma} \langle c_{s\bm{k} m \sigma}^{\dagger} c_{s\bm{k} m \sigma} \rangle$, where $N_{\bm k}$ is the number of grid points in the Brillouin zone. 
The results are obtained at $\lambda=0.4$ and for $t_1 = 0.45$ in Fig.~\ref{Fig:paramagnetic_Hall}(b) and for $t_1 =0.1$ in Fig.~\ref{Fig:paramagnetic_Hall}(c). 
The results are symmetric with respect to $n_{{\rm e}}=2$ (half filling) because of the particle-hole symmetry. 
In Fig.~\ref{Fig:paramagnetic_Hall}(b), $\sigma_{xy}^{\rm SH}$ is quantized at a nonzero integer $2$ at 1/4 and half fillings, where the system is insulating, as shown in Fig.~\ref{Fig:paramagnetic_band}(a). 
Meanwhile, $\sigma_{xy}^{\rm SH}$ is quantized at $-4$ only at half filling in Fig.~\ref{Fig:paramagnetic_Hall}(c), since the system becomes insulating only at half filling, as shown in Fig.~\ref{Fig:paramagnetic_band}(b). 
Note that $\sigma_{xy}^{\rm SH}$ shows a sharp change with a sign reversal for carrier doping to half filling, which indicates the possibility of controlling and switching the spin current. 

The nonzero spin Hall conductivity in the paramagnetic state is induced by the hidden site-dependent antisymmetric spin-orbit coupling. 
Indeed, an effective antisymmetric spin-orbit coupling is obtained by taking into account the effect of $t_1$ perturbatively; it is shown to be proportional to 
$\pm (t_1^2/\lambda) [\cos (\sqrt{3}/2 k_x)+ \cos(k_y/2) ]\sin (k_y) s_z$ for A ($+$) and B ($-$) sublattices (the detailed derivation will be shown elsewhere). 
Note that the $\bm{k}$ dependence reflects the threefold rotational symmetry. 
The situation is similar to the so-called Kane-Mele model, i.e., a single-band Hubbard model with imaginary hopping between next nearest-neighbor sites~\cite{Haldane_PhysRevLett.61.2015,Kane_PhysRevLett.95.226801}. 
Especially, among the different phases in Fig.~\ref{Fig:paramagnetic_Hall}(a), 
the origin of the quantum spin Hall insulator with $\sigma_{xy}^{{\rm SH}}=2$ is essentially the same as that in the Kane-Mele model.  
This means that our two-orbital model is considered to be an extension of the Kane-Mele model to multi-orbital cases. 
Indeed, our model exhibits richer behavior of the quantum spin Hall effect than the previous model.

\section{Summary and Concluding Remarks}
We have investigated the topological aspect of a two-orbital model on a honeycomb lattice. 
We have clarified the topological phase diagram at half filling, which includes three topologically-different insulators with different quantized values of the spin Hall conductivity. 
The results are discussed in terms of the hidden site-dependent antisymmetric spin-orbit coupling. 

Concerning the site-dependent antisymmetric spin-orbit coupling, it is interesting to consider the possibility of the linear magnetoelectric effect as discussed in Refs.~\cite{Yanase:JPSJ.83.014703,Hayami_PhysRevB.90.024432}. 
Recently, the authors showed that a global antisymmetric spin-orbit coupling is induced in the present model and the linear magnetoelectric effect appears when a long-range electronic order occurs in a way of breaking the global inversion and rotational symmetries~\cite{Hayami_PhysRevB.90.081115}. 
Similarly, it is possible to have the magnetoelectric effect even in the paramagnetic state once the rotational symmetry is broken by some perturbations, such as uniaxial pressure and defects.

\ack

S.H. is supported by Grant-in-Aid for JSPS Fellow. 
This work was supported by Grants-in-Aid for Scientific Research (No.~24340076), the Strategic Programs for Innovative Research (SPIRE), MEXT, and the Computational Materials Science Initiative (CMSI), Japan.

\section*{References}
\bibliographystyle{iopart-num}
\bibliography{ref}

\end{document}